\documentclass{elsart}

\usepackage{graphicx}
\usepackage{amssymb}

\begin{document}

\begin{frontmatter}

% Title, authors and addresses

% use the thanksref command within \title, \author or \address for footnotes;
% use the corauthref command within \author for corresponding author footnotes;
% use the ead command for the email address,
% and the form \ead[url] for the home page:
% \title{Title\thanksref{label1}}
% \thanks[label1]{}
% \author{Name\corauthref{cor1}\thanksref{label2}}
% \ead{email address}
% \ead[url]{home page}
% \thanks[label2]{}
% \corauth[cor1]{}
% \address{Address\thanksref{label3}}
% \thanks[label3]{}

\title{The 270 MeV deuteron beam polarimeter at the Nuclotron Internal Target Station}

% use optional labels to link authors explicitly to addresses:
% \author[label1,label2]{}
% \address[label1]
% \address[label2]{}
\author[a,k]{P.K.Kurilkin}, 
\author[a,k]{V.P.Ladygin}\footnote{vladygin@jinr.ru},
\author[b]{T.Uesaka},
\author[c]{K.Suda},
\author[a]{Yu.V.Gurchin},
\author[a]{A.Yu.Isupov},
\author[d]{K.Itoh},
\author[a,e]{M.Janek},
\author[a,f]{J.-T.Karachuk},
\author[b]{T.Kawabata},
\author[a]{A.N.Khrenov},
\author[a]{A.S.Kiselev},
\author[a]{V.A.Kizka},
\author[g]{J.Kliman}, 
\author[a,h]{V.A.Krasnov}, 
\author[a,h]{A.N.Livanov},  
\author[i]{Y.Maeda}, 
\author[a]{A.I.Malakhov}, 
\author[g]{V.Matousek}, 
\author[g]{M.Morhach},  
\author[a]{S.G.Reznikov}, 
\author[b]{S.Sakaguchi}, 
\author[b,j]{H.Sakai},  
\author[b]{Y.Sasamoto},  
\author[c]{K.Sekiguchi},  
\author[g]{I.Turzo}, 
\author[a,k]{T.A.Vasiliev}

\address[a]{Joint Institute for Nuclear Research, Dubna, Russia}
\address[b]{Center for Nuclear Study, University of Tokyo, Tokyo 113-0033, Japan}
\address[c]{RIKEN Nishina Center, Saitama, Japan}
\address[d]{Department of Physics, Saitama University, Saitama, Japan}
%\address[e]{P.J.\v Safarik University, Ko\v sice, Slovakia}
\address[e]{Physics Dept., University of \v Zilina, 010 26 \v Zilina, Slovakia}
\address[f]{Advanced Research Institute for Electrical Engineering, Bucharest, Romania}
\address[g]{Institute of Physics of Slovak Academy of Sciences, Bratislava, Slovakia}
\address[h]{Institute for  Nuclear Research, Moscow, Russia}
\address[i]{Kyushi University, Hakozaki, Japan}
\address[j]{University of Tokyo, Tokyo, Japan}
\address[k]{Moscow State Institute of Radio-engineering Electronics and Automation (Technical University), Moscow, Russia}

\begin{abstract}
% Text of abstract
A deuteron beam polarimeter has been constructed at the Internal Target Station 
at the Nuclotron of JINR. The polarimeter is based on spin-asymmetry measurements 
in the $d-p$ elastic scattering at large angles and the deuteron kinetic energy of 270 MeV. 
It allows to measure vector and tensor components of the deuteron beam polarization 
simultaneously. 
\end{abstract}


\end{frontmatter}

% main text
\section{Introduction}
\label{intro}
One of the directions of the nuclear spin physics at intermediate energies is the study of  the two- and three-nucleon forces spin structure in the deuteron induced reactions. Such investigations have been performed recently at 
RARF \cite{RIKEN,kimiko1,kimiko2,suda,kimiko_br1}, RCNP \cite{sag03},  KVI \cite{meh07,bina}  and IUCF \cite{cadman}.  High accuracy polarization data have been obtained. 
The experiments have been proposed at the Nuclotron \cite{uesaka} and RIBF \cite{kimiko-fb19} to study the energy dependence of the three-nucleon forces spin structure via measuring the
analyzing powers in $d-p$ elastic scattering. 

However, these studies require the high precision polarimetry to obtain reliable values of beam polarization.
On the one hand, since deuteron is a spin-1 particle, the polarimetry should have a capability to determine simultaneously both vector and tensor components of the beam polarization. Moreover, the effective analyzing powers of the polarimeter should be known with high precision to provide small systematic errors while determing the beam polarization components. 

The $d-p$ elastic scattering has been traditionally used for the tensor and vector polarimetry at intermediate
and high energies. 
It was demonstrated that the $d-p$ elastic scattering at forward angles has large vector $A_{y}$
 and tensor $A_{yy}$ analyzing powers at 1600 MeV \cite{ableev} and can be used for polarization 
analysis \cite{dpel_anl,dpel_1600}. Shortcoming of the measurements at forward angles is
that it requires a sophisticated equipment to identify the events. For example, in Ref.~\cite{ableev}, 
deuterons scattered at $\theta_{lab}^d=7.5^\circ$ 
from a hydrogen target were selected by the two-arm magnetic spectrometer ALPHA, from the deuterons of other reaction channels and other particles.
The same method has been applied recently at COSY using the ANKE spectrometer \cite{cosy}
at the initial deuteron energy of 1170 MeV and $4^\circ\le\theta_{lab}^d\le 10^\circ$ where the vector and tensor analyzing powers are large \cite{dpel_anl,dpel_1200}.

The $d-p$ elastic scattering {at large angles} ($\theta_{\rm cm}\ge 60^{\circ}$)
has been successfully used for the deuteron polarimetry at RIKEN at a few hundreds of MeV. This  reaction has several advantages as a beam-line
polarimetry over the others. Firstly, both the vector and tensor analyzing
powers for the reaction have large values \cite{RIKEN,kimiko1,kimiko2,dpel_395}. 
The values of the analyzing powers were obtained for the polarized deuteron beam, whose absolute polarization had been calibrated via the ${\rm ^{12}C}(d,\alpha){\rm ^{10}B^*[2^+]}$ reaction \cite{suda}. 
The determination precision of the deuteron beam polarization values using this method is better than 2\%.
%Secondly, a kinematical coincidence measurement of deuteron and proton with simple plastic scintillation counters suffices for event identification.
Secondly, to indentify the events, it is sufficient to carry out kinematical coincidence measurements of deuteron and proton by means of simple plastic scintillation counters.
 It is justified because of a small background event rate in comparison with the forward angles.

The goal of the present article is to report about a new polarimeter based 
on the asymmetry measurements in $d-p$ elastic scattering at 270 MeV at the Internal Target Station (ITS) \cite{ITS} of the  
Nuclotron (LHEP-JINR).  The paper is organized as follows. The details of the polarimeter construction are given in Section \ref{polarimeter}. The experimental and
data analysis procedures are described in Section \ref{experiment}. 
The details of the beam polarization evaluation and results
are given in Section \ref{sec:results}.  
The conclusions are drawn in the last Section.

%\appendix;
% appendix sections are then done as normal sections
% \appendix

\section{Polarimeter}
\label{polarimeter}

The efficient polarimetry can be provided even for rather low beam
intensity using a thin solid internal target inside the accelerator ring.
Due to a multi passage of the beam through the target point, the 
luminosity can be significantly increased by using the properly tuned
internal target trajectory. Therefore, this internal beam polarimeter
with a very thin target can have approximately the same performance as the external   
beam polarimeters.

The deuteron beam 
polarimeter was constructed at ITS in the Nuclotron ring. 
ITS consists of a spherical scattering chamber and a target sweeping system where six different targets can be mounted. 
The 10~$\mu$m~CH$_2$ film was used as a proton target for the polarimeter.
The details of ITS can be found in Ref.\cite{ITS}.

\begin{figure}[htbp]
 \centering
 \resizebox{10cm}{!}{\includegraphics{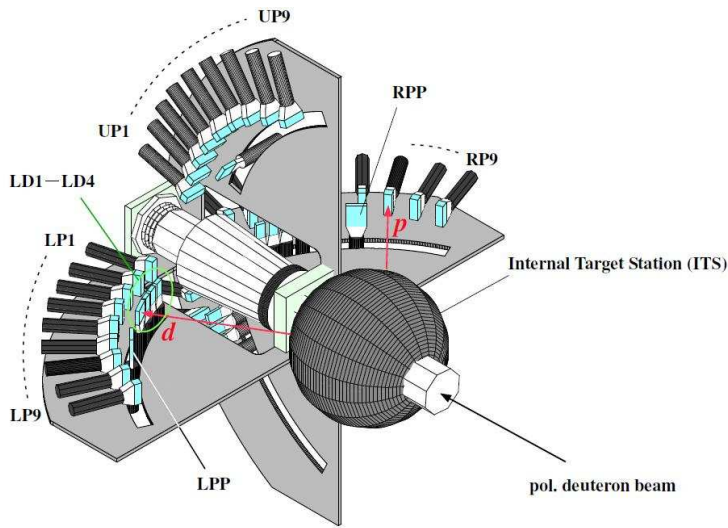}}
\caption{A schematic view of the polarimeter setup installed downstream the ITS. 
Plastic scintillation counters coupled to PMTs are placed to the left, right, up, and down the beam axis. \label{fig:setup}}
\end{figure}

A schematic view of the polarimeter is shown in Fig.\ref{fig:setup}. A detector support
with 43 mounted plastic scintillation counters is placed downstream the ITS. Each
plastic scintillation counter was coupled to a photo-multiplier tube Hamamatsu H7416MOD. 
Eight proton detectors were used for LEFT (LP1--8), as well as for RIGHT
(RP1--8) and UP (UP1--8), but due to space limitation -- only four for DOWN (DP1--4).
The proton detectors were placed 630 mm from the target. The angular span
of one proton detector was 2$^\circ$ in the laboratory frame, which corresponds to
$\sim$4$^\circ$ in the c.m. system.
Four deuteron detectors (LD1--4, RD1--4, DD1--4) were placed at scattering
angles of deuterons coinciding kinematicaly with the protons. On the other hand, one deuteron
detector (UD1) can cover the solid angle corresponding to DP1--4. 
The scattered deuterons and recoil protons were detected in kinematic coincidence
over the c.m. angular range of 65--135$^\circ$.

In addition, one pair of detectors (LPP and RPP) was placed to register two protons from quasi-elastic $p-p$ scattering at $\theta_{pp}$=90$^\circ$ in the c.m. in the horizontal plane. 
The deuteron and quasi-elastic
$p-p$ detectors were set in front of the proton detectors.

Sizes of the detectors and their setting angles for 270 MeV are listed in Table~\ref{tbl:detectors}.

\begin{table}[htbp]
\centering
\caption{Plastic scintillation counters used for the polarimeter. The setting angles of the detectors at 270 MeV
are shown in the last column. \label{tbl:detectors}}
\begin{tabular}{lcccc}
  \hline\hline
             & width & height  & thickness & LAB angle \\
             &  [mm]      & [mm]         &  [mm] & [deg] \\
  \hline
     proton detectors & 20 & 40/60  & 20 & 21.3, 26.1, 30.9, 35.8, \\
                      &    &        &    & 40.8, 45.0, 50.8, 55.9 \\
     deuteron detectors & 24 & 40 & 10 & 20.1, 22.7, 25.6 \\
                        & 50 & 40 & 10 & 29.3            \\
     quasifree $p-p$ detectors & 50 & 60 & 10 & 44.0 \\
  \hline\hline
\end{tabular}
\end{table}

For the internal beam polarimeter the luminosity 
depends both on the beam intensity and internal target trajectory. In this respect, the monitor reflecting
the number of the beam-target interactions, has to be used to measure the luminosity. 
On the other hand, the luminosity monitor has to be insensitive to the beam polarization. 

The quasi-elastic $p-p$ scattering yield at $\theta_{pp}$=90$^\circ$ in the c.m. in the horizontal plane meets the above requirements. On the one hand, it is defined by the number of beam-target collisions. On the other hand, it is independent of the beam polarization value, because the beam polarization axis in the Nuclotron ring has the vertical direction. 

\begin{figure}
  \centering
  \resizebox{10cm}{!}{\includegraphics{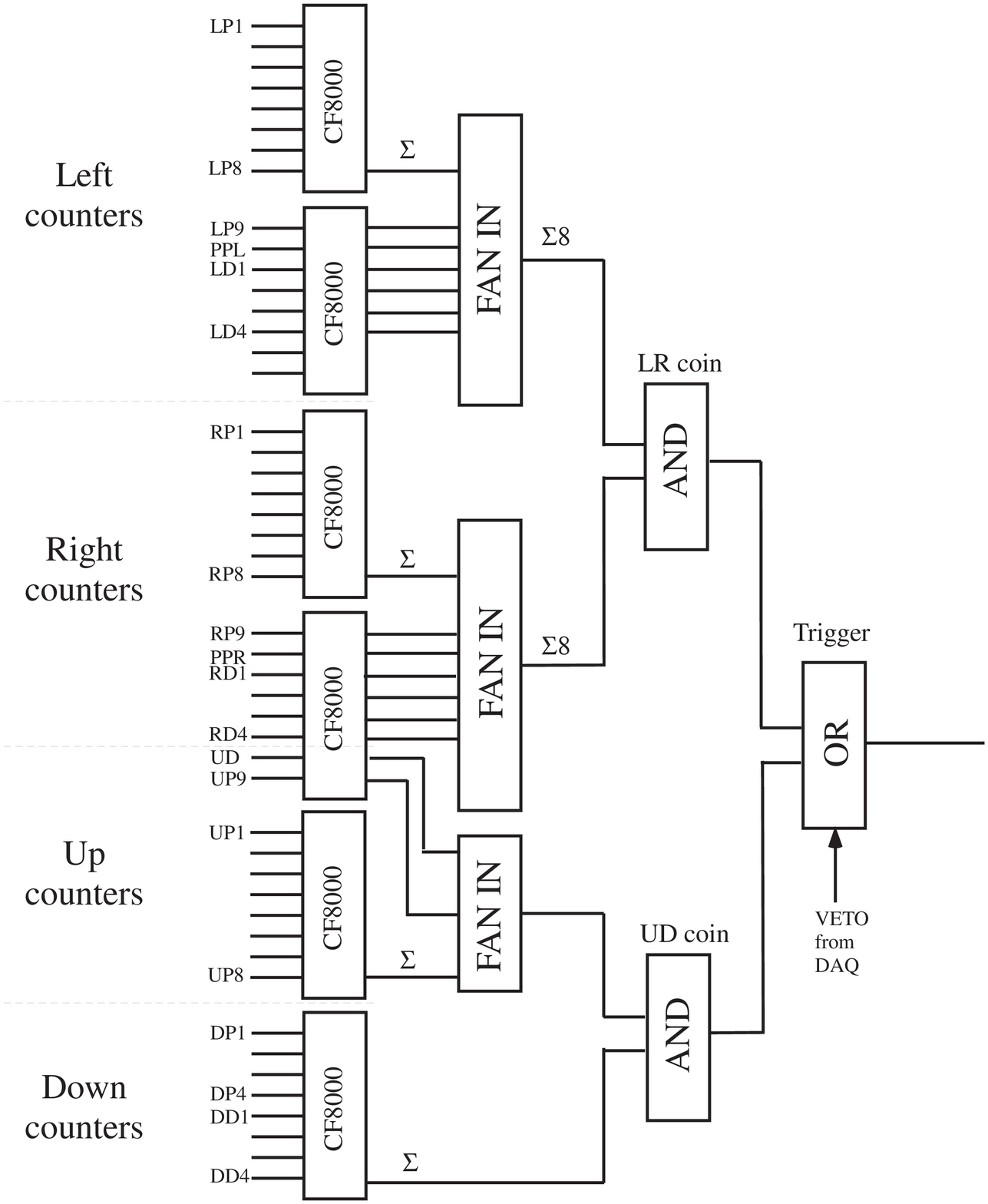}}
  \caption{Trigger logic for coincidences. \label{fig:circuit}}
\end{figure}

Anode signals from the photo-multiplier tubes were fed into the 8-channel
constant-fraction discriminators (CFD), ORTEC CF8000. CF8000 has one
analog output and three logic outputs for each channel. 
The analog signals from rear LEMO connectors were transmitted to Fast-Encoding
Readout ADCs (FERA), LeCroy 4300B, through cable delays. One of the logic
outputs from a rear ECL connector, was used to stop Time-to-FERA
Converter (TFC), LeCroy 4303. The other two logic outputs made a logical AND which corresponded to a coincidence between the
scattered deuteron and the recoiled proton. As shown in Fig.~\ref{fig:circuit},
the coincidences of any of LEFT detectors (LP1--8, LD1--4, PPL) and any of RIGHT detectors (RP1--8, RD1--4, PPR) are triggering the data taking. Coincidences between the UP and DOWN detectors were also used to make a trigger. The data were accumulated with a VME-CAMAC based
data-acquisition system \cite{OkamuraDAQ}.

\section{Experimental procedure}
\label{experiment}

The polarized deuteron beam was provided by the atomic-beam-type polarized ion source POLARIS
\cite{polaris}.  In the given experiment the data were taken for three spin-modes:
unpolarized, "2-6" and "3-5", which have theoretical maximum polarizations of  
$(p_z,p_{zz})$ =  $(0,0)$, $(1/3,1)$ and  $(1/3,-1)$, respectively. 
The spin modes were changed cyclically and spill-by-spill.  
The quantization axis was perpendicular to the beam-circulation plane of the Nuclotron. 

The typical beam intensity in the Nuclotron ring was 2--3$\times 10^7$ deuterons per spill with a duration of $\sim$1 s.
The repetition rate and orbit frequency at 270 MeV were 1/6 Hz and 0.583 MHz, respectively.
The 10~$\mu$m~CH$_2$ film was used as a proton target.
No measurements with the carbon target 
were made, because the background from the carbon content of the
$CH_2$ target was estimated as less than 1\% at 270 MeV \cite{RIKEN}.

\begin{figure}[htbp]
 \centering
 \resizebox{10cm}{!}{\includegraphics{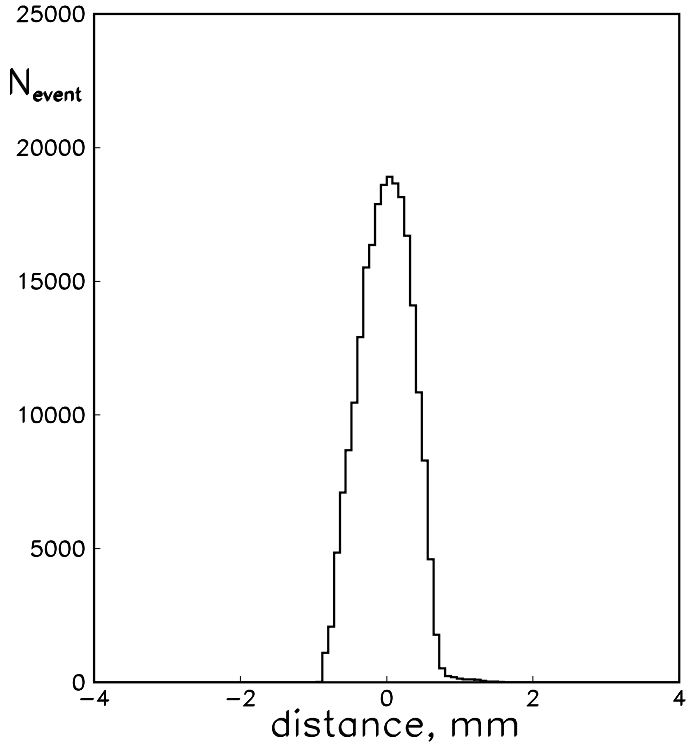}}
\caption{Distribution of the interaction point in mm. \label{fig:targetposition}}
\end{figure}

A signal from the target position monitor \cite{gurchin} was used to tune accelerator parameters to bring the interaction point close to the center of the ITS chamber. 
The signals from the monitor were also stored in raw data so that one can use the position information in off-line analysis. The distribution of the interaction point in mm is shown in Fig.\ref{fig:targetposition}. 
The uncertainty of the interaction point is estimated to be of $\pm 1$ mm. 
The details of the target position monitor  operation are described in Ref.~ \cite{gurchin}.

The $d-p$ elastic events were  selected by using the energy loss correlation and time-of-flight difference for the scattered deuteron and recoil proton. 

\begin{figure}[htbp]
 \centering
  \resizebox{10cm}{!}{\includegraphics{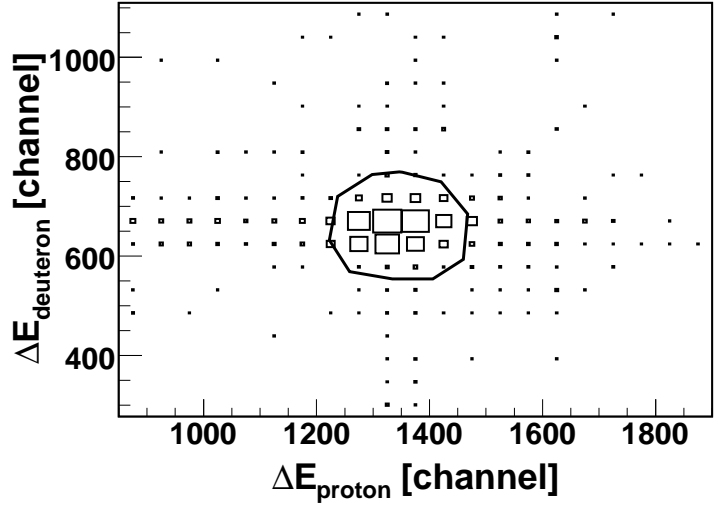}}
\caption{ The correlation  of the amplitude signals for one pair of the deuteron and proton detectors 
at 270 MeV. The solid line is a graphical cut to select the $d-p$  elastic events. \label{fig:elosscorr}}
\end{figure}

Fig.~\ref{fig:elosscorr} shows the energy loss correlation of proton and deuteron 
for the scattering angle of 75$^\circ$ in the c.m. 
One can see a prominent locus corresponding to the $d-p$ elastic events.
The solid line is a graphical cut for the selection of 
the $d-p$  elastic events. 

\begin{figure}[htbp]
 \centering
  \resizebox{10cm}{!}{\includegraphics{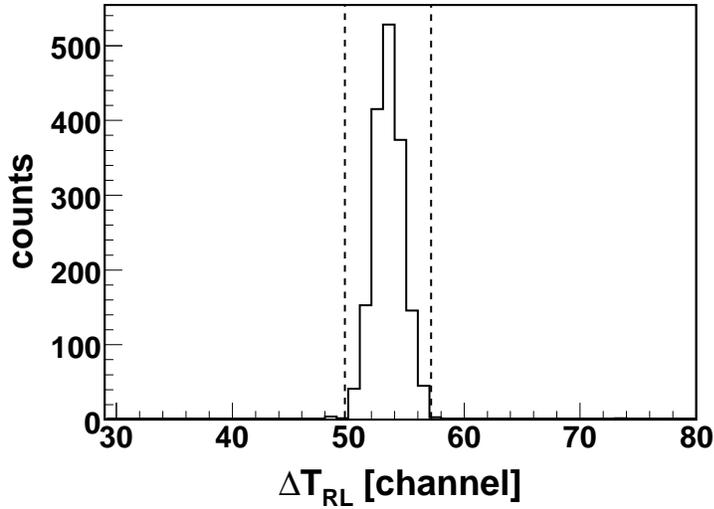}}
\caption{The time difference between the signals for deuteron and proton detectors 
at 270 MeV and a scattering angle of 75$^\circ$ in the c.m. 
\label{fig:timediff}}
\end{figure}

The final selection of the events 
used to deduce the spin-dependent asymmetry
was made by the time difference between the signals for the
conjugated deuteron and proton detectors.
Figure~\ref{fig:timediff} shows the time difference between the
signals for the deuteron and proton detectors at 
75$^\circ$ in the c.m. 
The dashed lines are the prompt timing windows to select $d-p$ elastic events.

\section{Deuteron beam polarization}
\label{sec:results}

\begin{figure}[htbp]
 \centering
  \resizebox{10cm}{!}{\includegraphics{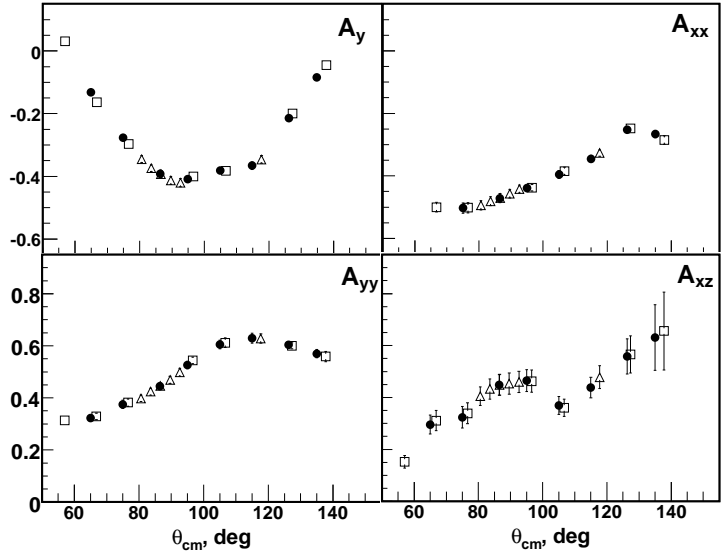}}
\caption{Analyzing powers $A_y$, $A_{yy}$, $A_{xx}$ and $A_{xz}$ of 
$d-p$ elastic scattering at 270 MeV as function of the scattering angle 
in the c.m. The open squares and triangles are the data from Refs. 
\cite{kimiko1} and \cite{kimiko2,suda}, respectively. 
The extrapolated values of the analyzing powers used to determine the deuteron beam
polarization, are shown by the solid circles. 
\label{fig:an_powers}}
\end{figure}

The beam polarization was determined by using asymmetry of the $d-p$ scattering yields and 
the known analyzing powers of the reaction \cite{kimiko1,suda}.
Data at several scattering angles:
75$^\circ$, 86.5$^\circ$, 95$^\circ$, 105$^\circ$, 115$^\circ$,
126.3$^\circ$  and 135$^\circ$,
were used  to increase the polarimeter figure of merit.
The values of the analyzing powers $A_y$, $A_{yy}$, $A_{xx}$ and $A_{xz}$ 
at these angles were obtained by the cubic spline interpolation of the
data taken from Refs.~\cite{kimiko1,suda}. 
The extrapolated values of the analyzing powers are shown in  Fig.\ref{fig:an_powers} 
by the solid circles and given in Table \ref{tbl:an_powers}. 
The open  squares and triangles represent 
the data used for the extrapolation procedure from Refs.~\cite{kimiko1} and 
\cite{kimiko2,suda}, respectively. The errors for the data from
Refs.~\cite{kimiko1,kimiko2,suda} were taken as quadrature root of the sum of squared  statistical and systematic errors.

\begin{table}[htbp]
\centering
\caption{The extrapolated values of the
analyzing powers  $A_y$, $A_{yy}$, $A_{xx}$ and $A_{xz}$ for $d-p$ elastic scattering
at 270 MeV used to determine the deuteron beam polarization values.
\label{tbl:an_powers}} 

\begin{tabular}{ccccccccc}
\hline
\hline
$\theta_{cm}$, [deg] & $A_y$ & $dA_y$ & $A_{yy}$  & $dA_{yy}$ 
 & $A_{xx}$ & $dA_{xx}$ &  $A_{xz}$  &   $dA_{xz}$ \\
\hline                  
65   &   -0.133 & 0.004 & 0.323 &  0.012 &    --  &   --   &  0.296 &  0.037 \\
75   &   -0.277 & 0.004 & 0.375 &  0.016 & -0.503 &  0.016 &  0.324 &  0.042 \\
86.5 &   -0.392 & 0.012 & 0.445 &  0.013 & -0.471 &  0.014 &  0.445 &  0.040 \\
95   &   -0.410 & 0.009 & 0.526 &  0.014 & -0.439 &  0.012 &  0.465 &  0.042 \\
105  &   -0.382 & 0.004 & 0.605 &  0.017 & -0.395 &  0.013 &  0.370 &  0.034 \\
115  &   -0.367 & 0.012 & 0.629 &  0.019 & -0.345 &  0.011 &  0.439 &  0.039 \\
126.3&   -0.216 & 0.007 & 0.604 &  0.011 & -0.253 &  0.008 &  0.558 &  0.067 \\
135  &   -0.084 & 0.005 & 0.570 &  0.016 & -0.266 &  0.012 &  0.631 &  0.126 \\
\hline  
\hline       
\end{tabular}
\end{table}

The $d-p$ scattering yields in "2-6" and "3-5" spin modes are normalized on the yield for the unpolarized mode after correction of the integrated beam intensity
and dead time of the data-taking system. For this purpose the absolute value
of beam intensity is not necessary and the relative value monitored by the quasi-
elastic $p-p$ scattering, is sufficient.

The normalized yields of the $d-p$ elastic events for the left ($L$), right ($R$), up ($U$) and down ($D$) scattering 
\begin{eqnarray}
\label{ds_deut1}
L &=& 1 +\frac{3}{2}p_y A_y  + \frac{1}{3} (2p_{xx}+p_{yy})A_{xx}
+\frac{1}{3}(2p_{yy}+p_{xx})A_{yy}\\
\label{ds_deut2}
R &=& 1 -\frac{3}{2}p_y A_y  + \frac{1}{3} (2p_{xx}+p_{yy})A_{xx}
+\frac{1}{3}(2p_{yy}+p_{xx})A_{yy}\\
\label{ds_deut3}
U &=& 1   + \frac{1}{3} (2p_{yy}+p_{xx})A_{xx}+\frac{1}{3}(2p_{xx}+p_{yy})A_{yy}\\
\label{ds_deut4}
D &=& 1  + \frac{1}{3} (2p_{yy}+p_{xx})A_{xx}
+\frac{1}{3}(2p_{xx}+p_{yy})A_{yy}
\end{eqnarray}
were used to determine the beam polarization. Here,
$A_y$, $A_{yy}$ and $A_{xx}$  denote the deuteron analyzing powers, 
$p_y$, $p_{yy}$ and $p_{xx}$  are
the corresponding components of the beam polarization \cite{ohlsen}.

The $L$, $R$, $U$ and $D$ yields of $d-p$ elastic scattering events at the scattering angles $\theta_{cm}\ge 105^\circ$; while only the $L$, $R$ and $D$ yields at the 
angles of 75$^\circ$, 86.5$^\circ$ and 95$^\circ$
were used to determine beam polarization values. 

\begin{figure}[htbp]
 \centering
  \resizebox{10cm}{!}{\includegraphics{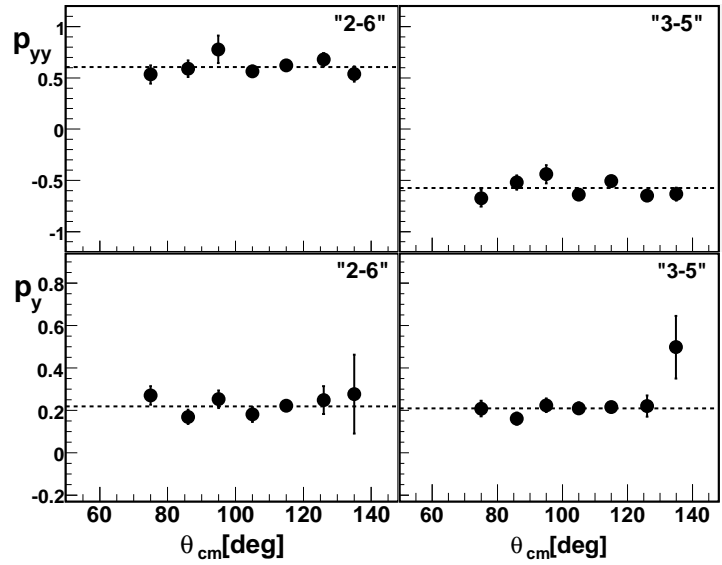}}
\caption{Tensor $p_{yy}$ and vector $p_y$ polarizations of the beam for  "2-6" and "3-5" spin modes of 
POLARIS \cite{polaris} as function of the deuteron scattering angle in 
the c.m.
\label{fig:py_pyy_cm}}
\end{figure}

Fig.~\ref{fig:py_pyy_cm} demonstrates the values of the tensor $p_{yy}$ and vector $p_y$ polarizations of the beam for 
"2-6" and "3-5" spin modes of 
POLARIS \cite{polaris} as function of the deuteron
scattering angle in 
the c.m. 
The error bars include both statistical and systematic errors which are related with the uncertainties in the values of analyzing powers. 
One can see a good agreement of the polarization values obtained at different scattering angles in the c.m. 
The dashed lines in Fig.~\ref{fig:py_pyy_cm} represent the beam polarization values  averaged over all the scattering angles.

\begin{figure}[htbp]
 \centering
  \resizebox{10cm}{!}{\includegraphics{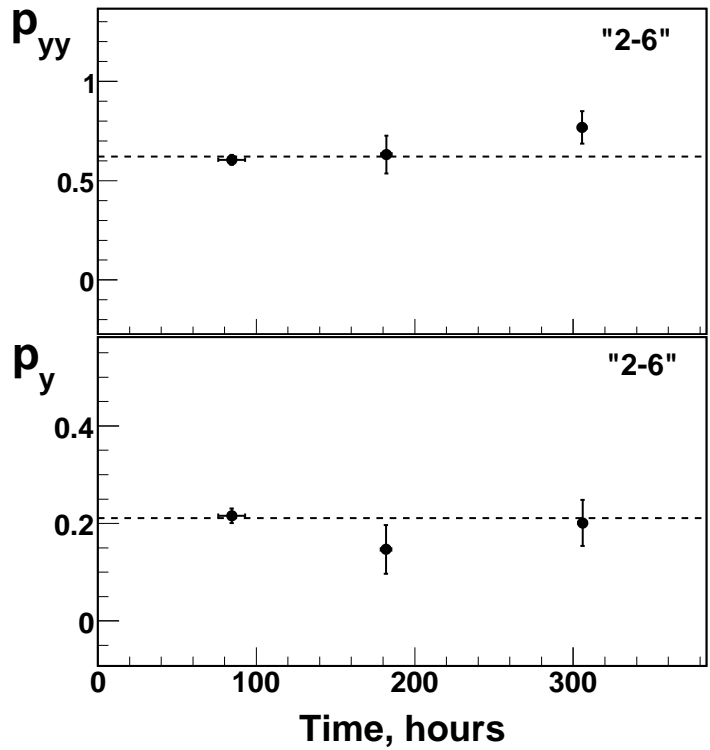}}
\caption{Tensor $p_{yy}$ and vector $p_y$ polarizations of the beam for the spin mode "2-6" of 
POLARIS \cite{polaris} versus the measuring time in hours. 
\label{fig:py_pyy_t26}}
\end{figure}

\begin{figure}[htbp]
 \centering
  \resizebox{10cm}{!}{\includegraphics{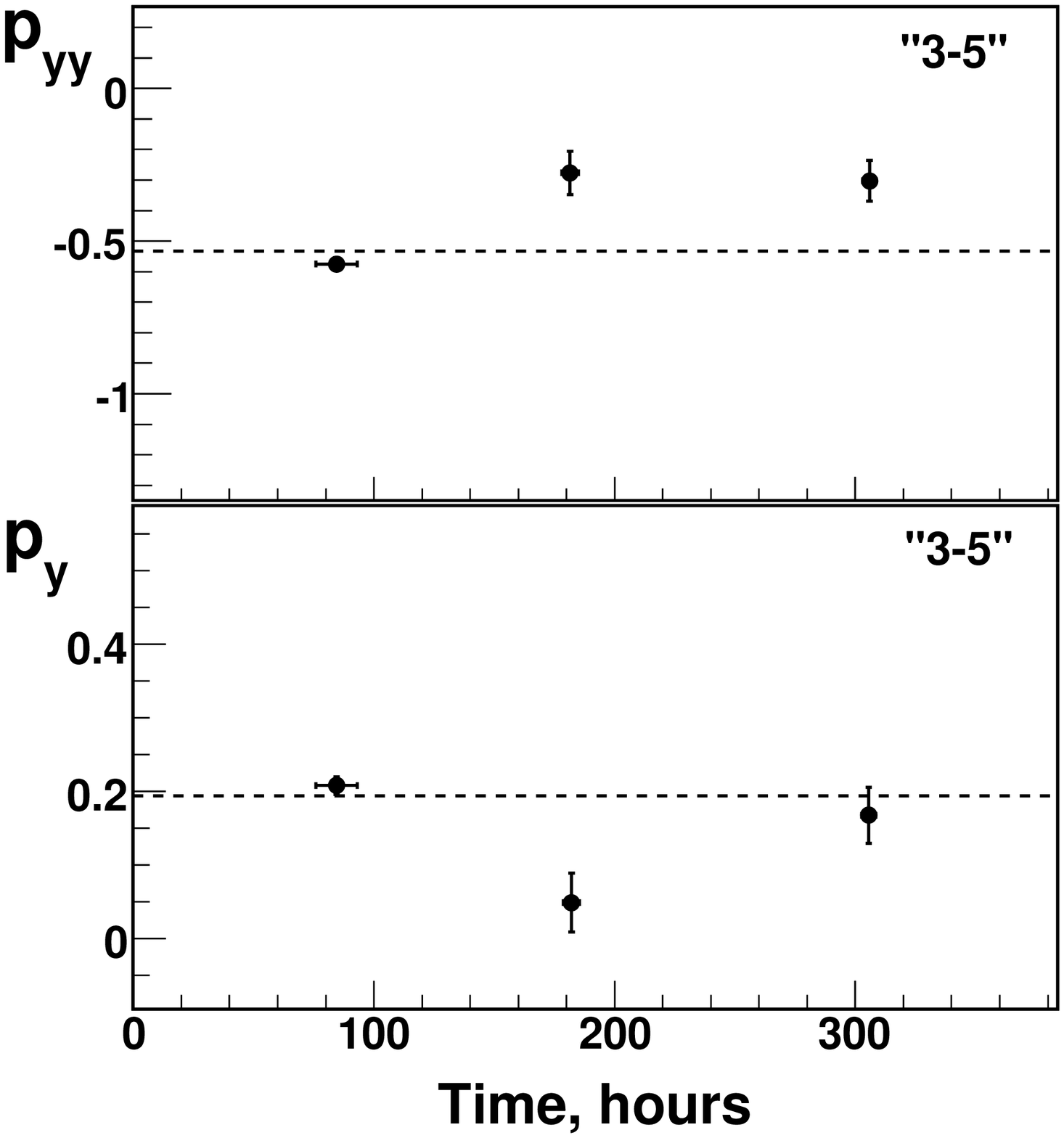}}
\caption{Tensor $p_{yy}$ and vector $p_y$ polarizations of the beam for the spin mode "3-5" of POLARIS \cite{polaris} versus the measurement time in hours. 
\label{fig:py_pyy_t35}}
\end{figure}

The beam polarization was measured at 270 MeV several times  
during the experiment studying the analyzing
powers in $d-p$ elastic scattering at high energies \cite{dubna880}. 
Figs.\ref{fig:py_pyy_t26} and \ref{fig:py_pyy_t35} illustrate the polarization values for the spin modes "2-6" and "3-5" of 
POLARIS \cite{polaris}, respectively, as functions of the measuring time in hours.
One can see good time stability of the beam polarization values during the experiment ($\sim$220 hours).

Table~\ref{tbl:polarization} gives the values of the tensor $p_{yy}$ and vector $p_y$ polarizations of the beam for 
"2-6" and "3-5" spin modes of 
POLARIS \cite{polaris} averaged over the duration of the experiment.
Both statistical and systematic errors due to the uncertainties in the values of the analyzing powers are shown in the above Table. The systematic errors do not exceed
1-2\% for the both vector and tensor polarizations of the beam. 
The angle between the beam direction and polarization axis was found to be 
$\beta=-90.3\pm 1.2^\circ$, namely, the polarization axis was perpendicular to the
plane containing the mean beam orbit in the accelerator.

\begin{table}
\centering
\caption{The averaged values of the vector $p_y$ and tensor $p_{yy}$  beam polarizations 
for the
"2-6" and "3-5" spin modes of POLARIS \cite{polaris}.\label{tbl:polarization}}  
 
\begin{tabular}{ccccccc}
\hline\hline
Spin mode & $P_y$ & $\Delta P_y^{stat}$ &  $\Delta P_y^{sys}$ &
$P_{yy}$ & $\Delta P_{yy}^{stat}$ &$\Delta P_{yy}^{sys}$ \\
\hline\
"2-6" & 0.210 & 0.013 & 0.002 &  0.619 & 0.022 & 0.005 \\
"3-5" & 0.193 & 0.011 & 0.002 & $-$0.532 & 0.018 & 0.004\\
\hline\hline         
\end{tabular}
\end{table}

The performance of the polarimeter is expressed in terms of the
figure of merit, $\cal F$. It allows one to evaluate the counting rate
$N_{inc}$, needed for the desired beam polarization accuracy $\Delta P$:
\begin{eqnarray}
\label{deltap}
\Delta P \sim \frac {\sqrt{2}}{{\cal F}\sqrt{N_{inc}}},
\end{eqnarray}

$\cal F$
is a function of efficiency $\epsilon$ and analyzing power $A$.
It is defined as
\begin{eqnarray}
\label{fmerit}
{\cal F}^2 = \int \epsilon\cdot A^2 d\Omega,
\end{eqnarray}
where $\Omega$ is the solid angle, and integration of (\ref{fmerit}) is curried out over the
angular domain of the polarimeter, efficiency $\epsilon= N_{ev}/N_{inc}$,  where $N_{ev}$ is the number of useful events detected and
$N_{inc}$ is the number of incident particles, $A$ is the analyzing power. 

In our case the number of incident particles $N_{inc}$ has been evaluated  from the number 
of $p-p$ quasi-elastic events at 90$^\circ$ in c.m. detected within the 
solid angle of $7.56\cdot 10^{-3}$ sr.  
The estimation of figures of merit was done as a sum over all the scintillation 
counters for $d-p$ elastic events detection for the scattering angles of 75$^\circ$-- 135$^\circ$ in c.m. Figures of merit ${\cal F}_y$,
${\cal F}_{yy}$ and ${\cal F}_{xx}$ were estimated as $\sim$1.6$\cdot$10$^{-4}$,
$\sim$2.1$\cdot$10$^{-4}$  and  $\sim$0.6$\cdot$10$^{-4}$, respectively.

These values are comparable with the figures of merit for the deuteron polarimeter used
at the extracted beam at RIKEN \cite{suda}.
However, in contrast with the RIKEN polarimeter \cite{suda} the large figures of merit values were obtained with a very thin CH$_2$ target 10 $\mu$m thick using the internal beam.  Estimated from $p-p$ quasi-elastic scattering at 
90$^\circ$ in c.m. the averaged luminosity was $\cal L$ $\sim$ 5.2$\cdot$ 10$^{28}$ cm$^{-2}$ per beam spill. 
This value corresponds to the CH$_2$ target effective length of
$\sim $ 5.2$\cdot$10$^{21}$ g/cm$^2$ (0.6~mm) for the external beam case.

The performance of the polarimeter can be increased by the 
extention of the detection system to smaller scattering angles in c.m.,
since in this region the cross section of $d-p$ elastic scattering is larger
and the analyzing powers have non-zero values. 
The detection of $d-p$ elastic scattering at small angles in c.m. is especially important
at higher energies since the cross section decreases while energy increasing.

The current polarimeter can be used in a wide deuteron energy range. On the other hand, 
the current setup was used to measure analyzing powers in $d-p$ elastic scattering
at higher energies \cite{dubna880} to study three-nucleon correlations.

\section{Conclusions}
\label{conclud}

\begin{sloppypar}

The new polarimeter based on $d-p$ elastic scattering has been constructed at the internal beam of the Nuclotron.
The measurements of the deuteron beam polarization have been performed at 270 MeV and at this energy there are already precise data on the $d-p$ elastic scattering analyzing powers in Refs.\cite{RIKEN,kimiko1,kimiko2,suda}. 

The detection at several scattering angles was used to compensate low intensity of the polarized beam from POLARIS \cite{polaris}. However, 
the systematic error due to the uncertainties in the values of the analyzing powers 
does not exceed 1-2\% for the both vector and tensor beam polarizations.

The figures of merit achieved are comparable with the ones for the  external beam polarimeter at RIKEN  \cite{suda} since the internal beam of the Nuclotron performs multi passages  through the thin CH$_2$ target.

This polarimeter can be used in a wide deuteron energy range \cite{dpel_395,dubna880}.
The future upgrade of the polarized ion source at the LHEP-JINR Accelerator Complex 
\cite{new_PIS} will greatly 
enhance the capability of the current setup to measure spin-observables of the $d-p$ scattering and perform the polarimeter calibration in a wide energy range.

The authors are grateful to the Nuclotron accelerator  and POLARIS groups.
They thank L.S.~Azhgirey, 
Yu.S.~Anisimov,  E.~Ayush, A.F.~Elishev, V.I.~Ivanov, L.V.~Karnjushina,  Z.P.~Kuznezova, 
A.P.~Laricheva, A.G.~Litvinenko, V.G.~Perevozchikov, V.M.~Slepnev,  Yu.V.~Zanevsky and 
V.N.~Zhmyrov  for their help during the preparation and performance of the 
experiment.
%The author express their appreciation to Yu.K.~Pilipenko and L.S.~Zolin for the
%measurements of the polarization at low energy polarimeter. 
The investigation has been  partly supported 
by the Grant-in-Aid for Scientific Research (Grant No. 14740151) of 
the Ministry of Education, Culture, Sports, Science, and Technology of Japan;
by the Russian Foundation for 
Fundamental Research (Grants No. 07-02-00102-a and 10-02-00087-a); by the Grant Agency
for Science at the Ministry of Education of the Slovak Republic 
(Grant  No. 1/4010/07) and by a Special program of the Ministry of Education and Science of the Russian Federation (Grant RNP2.1.1.2512). 

\end{sloppypar}

% Bibliographic references with the natbib package:
% Parenthetical: \citep{Bai92} produces (Bailyn 1992).
% Textual: \citet{Bai95} produces Bailyn et al. (1995).
% An affix and part of a reference:
%   \citep[e.g.][Ch. 2]{Bar76}
%   produces (e.g. Barnes et al. 1976, Ch. 2).

\end{document}